\documentclass{aa}
\usepackage{graphicx}

\begin{document} 

\title{Joint cosmological parameters forecast
from CFHTLS-cosmic shear and CMB data}

\author{I. Tereno 
          \inst{1,3}
          \and
	  O. Dor\'e\inst{2} \and
          L. van Waerbeke\inst{1}
	  \and
	  Y. Mellier\inst{1,4}
          }

\offprints{Ismael Tereno (tereno@iap.fr) \\ }

\institute{Institut d'Astrophysique de Paris,
	   98 bis, boulevard Arago, F-75014 Paris, France
         \and
	    Department of Astrophysical Sciences, Princeton University,
	    Princeton NJ 08544, USA
	     \and 
             Departamento de F\'{i}sica, Universidade de Lisboa, 
	     1749-016 Lisboa, Portugal
	    \and 
            Observatoire de Paris, LERMA, 61 avenue de 
               l'Observatoire, F-75014 Paris, France
	     }	     

\date{}

\abstract{
We present a prospective analysis of a combined
cosmic shear and cosmic microwave background data set,
focusing on a Canada France Hawaii Telescope Legacy Survey (CFHTLS) type
lensing survey and the current WMAP-1 year and CBI data.
We investigate the parameter degeneracies and error estimates of a 
seven parameters model, for the lensing alone as well as for the combined 
experiments. The analysis is performed using a Monte
Carlo Markov Chain calculation, allowing for a more realistic estimate
of errors and degeneracies than a Fisher matrix approach.
After a detailed discussion of the relevant statistical
techniques, the set of
the most relevant 2 and 3-dimensional lensing contours are given.
It is shown that the
combined cosmic shear and CMB is particularly efficient to break some
parameter degeneracies.
The principal components
directions are computed and it is found that the most orthogonal contours
between the two experiments
are for the parameter pairs $(\Omega_m,\sigma_8)$, $(h,n_s)$
and $(n_s,\alpha_s)$, where $n_s$ and $\alpha_s$ are the slope of the
primordial mass power spectrum and the running spectral index respectively.
It is shown that
an improvement of a factor $2$ is expected on the running spectral index
from the combined data sets.
Forecasts for 
error improvements from a wide field space telescope lensing survey
are also given.
}

\authorrunning{Tereno et al}


\maketitle

\section{Introduction}

The Canada-France-Hawaii Telescope Legacy Survey
\footnote{http://www.cfht.hawaii.edu/Science/CFHTLS/} (CFHTLS) 
 is a long term wide field imaging project that started in early 2003
 and should be completed by 2008.
   The French and Canadian astronomical communities will 
 spend about 500 CFHT nights to carry out imaging surveys
 with the new Megaprime/Megacam instrument recently mounted at the 
   CFHT prime focus.
  About 160 nights will focus on the "CFHTLS-Wide" survey
that will cover 170 deg$^2$, spread over 3 uncorrelated patches of 
 $7^o \times 7^o$ each,  in $u^*,g',r',i',z'$ bands, with typical exposure
  times of about one hour per filter.  The "CFHTLS-Wide" survey design 
    and observing strategy are similar 
  to the {\sc Virmos}-Descart cosmic shear survey
  \footnote{http://terapix.iap.fr/cplt/oldSite/Descart/}
  but it  
 will have a sky coverage 20 times larger.  It is widely 
   seen as a typical second generation cosmic shear survey.

The exploration of weak gravitational distortion produced by the large scale 
structures of the universe over field of views as large as "CFHTLS-Wide" 
 has an enormous potential for cosmology.  Past experiences based on 
 first generations cosmic shear surveys (see for example reviews in 
 \cite{VWM03,REFRE03}) have demonstrated they can 
 constrain the dark matter properties ($\sigma_8$, $\Omega_m$ and the shape 
 of the dark matter power spectrum) from a careful investigation of 
 the ellipticity induced by weak gravitational shear on distant galaxies. 
 For example, the  most recent cosmic shear results from the 
   {\sc Virmos}-Descart survey (\cite{VWMH04}) lead to
  the conservative limits $\sigma_8 =0.85 \pm0.15$ (99\% C.L.) and
  $\Omega_m=0.3\pm 0.15$ (99\% C.L.), which means an expected
accuracy of
  $\approx$1-3\% can be expected with the "CFHTLS-Wide" for the same set of
  cosmological parameters.  The CFHTLS-Wide will also
  explore a broader wavenumber range (10$^5$-10$^2$) than 
  {\sc Virmos}-Descart and will extend to  linear scale, that 
   will considerably ease cosmological interpretation of weak lensing data.
 The second generation surveys will therefore permit to   
   investigate more thoroughly different cosmological models, taking into
    account a broad range of cosmological parameters. For instance,
the CFHTLS as a probe dark energy evolution was stressed out by 
\cite{VWBENAB03}.

The full scientific outcome of
the cosmic shear data from the "CFHTLS-Wide" will only be complete
with a joint analysis with other data sets, like 
Type Ia Supernovae, galaxy redshift surveys, Lyman-alpha forest, 
or CMB observations.  \cite{CONTALDHOEK03} have used 
the Red Cluster Sequence (RCS)  cosmic shear survey together with the 
CMB data. It was shown that the $\Omega_m,\sigma_8$ degeneracies
for lensing and CMB are nearly orthogonal, which makes this set
of parameters particularly relevant for such combined analysis
(\cite{VW02}).
The search for orthogonal parameter
degeneracies between different observations
is one of the most important aspects of parameters measurements.
  
  $\,$\cite{ISHAK03} recently argued that joint CMB-cosmic shear 
  surveys provide an optimal data set to explore the amplitude of 
  the running spectral index and probe inflation models. They used a 
  Fisher-Matrix analysis on  WMAP+ACBAR+CBI and a "reference 
  survey". Their simulated survey  
  covers about 400 deg$^2$ with a depth corresponding to
    a galaxy number density of lensed 
  sources of about 60 arc-min$^{-2}$, and they restricted their 
  analysis to 3000$>l>$20. They found that several parameters 
  can be significantly improved (like $\sigma_8$, $\Omega_mh^2$, 
 $\Omega_{\Lambda}$)  and in particular that both the spectral 
 index $n_s$ and the running 
 spectral index $\alpha_s$ errors are reduced by a factor of 2. 
Their encouraging results show that joint CMB and weak lensing data 
  may provide interesting insights on inflation models.
Here, we investigate the 2-dimensional structure of
the parameter degeneracies between
lensing and CMB data sets, and look for the
expectation with a "CFHTLS-Wide"-like 
    survey design.  

 To explore the smaller scales probed by the "CFHTLS-Wide", which 
will provide cosmic shear information down to 20 arc-seconds, 
  it is preferable to avoid prior assumptions
 regarding the Gaussian nature of the underlying distribution, and to
 discard a Fisher matrix analysis. We used in this work the so-called
Markov Chain Monte Carlo (MCMC) method.
    The MCMC computing time linearly scales with the number 
  of parameters  
 and eases the exploration of a large sample of parameters and a broad 
    range of values for each. 
   \cite{CONTALDHOEK03} already used this approach with the RCS survey 
    to map the 
  $\Omega_m,\sigma_8$ parameter space, but marginalised 
 over a small set of cosmological parameters. 
 The goal of this present work  is  to  map the parameter space that 
 describes cosmological models in order to extract series 
 of parameter combinations that would minimise intersections of
 CMB and cosmic shear degeneracy tracks. 
Compared to the Fisher-Matrix approach which produces
 ellipses only, MCMC provides more details of the parameter
 space and eventually a more realistic estimate of 
  error improvements of the joint analyses.

  The paper is organised as follows: Section 2 introduces the gravitational
  lensing and defines the cosmic shear fiducial data used and the parameter
  space investigated. Section 3
  gives the details of our MCMC calculations, limitations and convergence
  criteria. Section 4 shows the MCMC results from the cosmic shear alone,
  assuming a lensing survey similar to the CFHTLS. In Section 5 we present the
  results of the parameter degeneracies analysis on the combined
  cosmic shear and cosmic microwave background observations.
  The assumptions made and the results obtained are discussed in section 6
  and we conclude in section 7.

\section{Cosmic shear and Cosmological parameters}

Propagation of galaxy light beams across
large scale mass inhomogeneities produces distorted and (de)magnified
galaxy images (for reviews see \cite{ME99,BS01,REFRE03,VWM03}).
The gravitational lensing 
magnification and shear are described by the
amplification matrix which involves second order derivatives of the 
projected gravitational
potential $\varphi$ : the shear, $\gamma$, and the
convergence, $\kappa$,
\begin{equation}
\kappa={1\over 2}\left(\varphi_{,11}+\varphi_{,22}\right)\ ; \ \
\gamma_1={1\over 2}\left(\varphi_{,11}-\varphi_{,22}\right)\ ; \ \
\gamma_2=\varphi_{,12}.
\end{equation}
The cosmic shear may be derived from the ellipticity of the galaxies. 
In the weak lensing  approximation, the observed 
 ellipticity of a galaxy is related to its
intrinsic ellipticity and to the shear by,
\begin{equation}
\label{wl}
\epsilon_i=\epsilon_{is}+\gamma \ ,
\end{equation}
where $\epsilon_i=(\epsilon_{xi},\epsilon_{yi})$ and 
$\gamma=(\gamma_1,\gamma_2)$.
Several correlation functions and 2-point statistics of the shear may be
defined. In this work we will use the shear variance
in a top-hat window of radius $\theta$ (Eq. (\ref{varshear})), although all 2-points
 statistics are equivalent
and provide similar results (they are all linear combinations of the others).
Let us now relate this quantity to cosmology.

We parameterize cosmological models with a set of 13 
parameters,
$(\omega_b,\omega_c,\Omega_v,f_\nu,h;\sigma_8,n_s,\alpha_s,r,n_t;\tau;w;
z_s)$.
Each model defines a point in the high-dimensional space where a value of 
the likelihood of the model with respect to the data may be calculated.
We use the CAMB software (\cite{LEWCHALAS00}) to compute the dark
matter power spectrum and transfer function. The parameters are defined
as follow:

\begin{itemize}
\item $(A_s,n_s,\alpha_s)$ - parameterize the primordial scalar power spectrum: 
\begin{equation}
\label{pkpri}
P(k)=A_s{\left(k\over{k_{0s}}\right)}^{n_s(k_{0s})-1+{1\over 2}\alpha_s \ln({k\over k_{0s}})},
\end{equation}
which is normalized with $A_s=P(k=0.05h{\rm Mpc}^{-1})$ and has a 
scale-dependent tilt.

\item $(A_s,r,n_t)$ - parameterize the primordial tensor power spectrum:
\begin{equation}
P(k)=rA_s{\left(k\over{k_{0t}}\right)}^{n_t},
\end{equation}
with $k_{0t}=0.002h{\rm Mpc}^{-1}$.

\item $(\omega_b,\omega_c,\Omega_v,f_\nu)$ - the matter budget consisting of baryons, cold dark
matter, dark energy and neutrinos, with $\omega_b=\Omega_b\,h^2$,
$\omega_c=\Omega_c\,h^2$.

\item $(h)$ - the hubble parameter. The spatial curvature is not a free parameter :
$\Omega_k=1-\Omega_v-(\omega_c+\omega_b)/h^2.$

\item $(w)$ - dark energy equation of state parameter.

\item $(\tau)$ - reionization optical depth.

\item $(\sigma_8)$ - Once the 3-dimensional power spectrum of the
density fluctuations is derived, it is renormalized according to the model value
of $\sigma_8$, losing the memory of the original value of $A_s$.

\end{itemize}
The non-linear evolution of the power spectrum is then 
evaluated using the {\sc halofit}  prescription of \cite{SMITHetal03}.
The power spectrum of the gravitational convergence 
  is derived by integrating the dark matter power spectrum 
 along the line-of-sight from the observer up to the
 radial coordinate of the 
 horizon $\chi_H$:
\begin{equation}
\label{p2d}
P_\kappa(k)={9\over 4}\Omega_m^2\int_0^{\chi_H} {\rm d}\chi{g^2(\chi) \over a^2(\chi)}
P_{3D}\left({k\over f_K(\chi)}; \chi\right) \ ,
\end{equation}
where $k$ is the 2-dimensional wave vector perpendicular
to the line-of-sight.  $f_K(\chi)$ is the comoving angular diameter
distance to a coordinate $\chi$, and $g(\chi)$ is given by
\begin{equation}
g(\chi)=\int_\chi^{\chi_H}{\rm d} \chi' p(\chi') 
{f_K(\chi'-\chi)\over f_K(\chi')},
\end{equation}
where $p(\chi(z))$ is the source redshift distribution, which depends
on the last item in our cosmological parameters set:
\begin{itemize}

\item $(z_s)$ - the redshift of the sources.

\end{itemize}
For simplicity, we assume that all the sources are at a single
redshift, $z_s$, but the generalisation to a broad redshift distribution
is straightforward.

From the power spectrum of the gravitational convergence,  
the top-hat shear variance inside a circle of radius $\theta$ 
can be computed:
\begin{equation}
\label{varshear}
\langle \gamma^2(\theta)\rangle={2\over \pi\theta^2}
\int_0^\infty~{{\rm d}k\over k}
P_\kappa(k) [J_1(k\theta)]^2.
\end{equation}
\begin{table}
\caption{Cosmic shear : fiducial cosmological model.}
\label{fiduc}
\begin{center}
\begin{tabular}{lll}
\hline \hline
 $\omega_b=0.022$ & $\omega_c=0.114$ & $\Omega_v=0.73$ \\
 $f_\nu=0$ & $\Omega_k=0$ & $h=0.71$ \\
 $n_s(k_{0s})=0.93$ & $\alpha_s(k_{0s})=-0.04$ & $r=0$ \\
 $w=-1$ & $\tau=0.17$ &\\
 $\sigma_8=0.9$ & $z_x=0.8$ & \\
 \hline \hline
\end{tabular}
\end{center}
\end{table}
For each cosmological model generated by the Monte Carlo simulation
described in the next section, 
$\langle \gamma^2(\theta)\rangle$ is computed and
compared to a fiducial data model, $\langle \gamma^2_{\;\rm data}
(\theta)\rangle$ that corresponds to the fiducial cosmological model
of Table \ref{fiduc}. This is taken to be
the best fit $\Lambda CDM$, flat, no neutrinos, no gravitational waves,
running spectral index model, derived from the 
first year WMAP data (\cite{SPERGEL03}). The sources are placed at
$z_s=0.8$.
The likelihood of each model is given by
\begin{equation}
\label{like}
-2\,{\cal L} = \left(\langle \gamma^2\rangle_i-\langle \gamma^2_{\;\rm data}
\rangle_i\right) C_{\rm data\, ij}^{-1}\left(\langle \gamma^2\rangle_{\,j}
-\langle \gamma^2_{\;\rm data}\rangle_j\right),
\end{equation}
where $C_{\rm data\, ij}$ is the covariance matrix
\begin{equation}
\label{covar}
C_{\rm data\, ij}=\langle\left(\gamma_i^2-\gamma_{{\rm data}~i}^2\right)
\left(\gamma_j^2-\gamma_{{\rm data}~j}^2\right)\rangle.
\end{equation}
%

The covariance matrix of a shear dispersion distribution is analytically
derived in \cite{SCHNEI02}. The shear dispersion at a given
 angular scale $\theta$ may be expressed as an integral of the two-point
correlation of galaxy ellipticities, $\xi_+$, estimated in
bins of angular separation $\Delta\vartheta$ at the center position 
 $\vartheta_i$, evaluated with a window function $S_+(\vartheta_i/\theta)$.
The covariance of the shear dispersion involves the four-point
ellipticities correlations.
Eq. (\ref{wl}) shows they depend on the
intrinsic ellipticity correlations, and therefore on 
  the intrinsic ellipticity
dispersion $\sigma_{\epsilon}$, and on the shear correlations
for each model. 
The shear field is assumed to be Gaussian, so that the
4-point function can be factorised as a sum over products
of 2-point functions.
 It is assumed that the covariance matrix depends
solely on the fiducial cosmological model.
Assuming a  
 connected single field of solid angle $A$ and a
mean density $n_g$ of source galaxies with an intrinsic
ellipticity dispersion of $\sigma_\epsilon$, the ensemble average of the
covariance matrix, ${\rm Cov}(\mathcal S;\theta_1,\theta_2)$ of the estimator
$\mathcal S$ of $\langle \gamma^2(\theta)\rangle$ is,
{\small{
\begin{eqnarray}
\label{bigcov}
\nonumber
{\rm Cov}(\mathcal S)={{\sigma^4_{\epsilon}}\over{2\pi\,A\,n_g^2}} 
\int_0^{2\rm min (\theta_1,\theta_2)}{d\vartheta\,\vartheta\over \theta_1^2\theta_2^2}
\left[S_+\left({\vartheta \over \theta_1}\right)S_+\left({\vartheta \over \theta_2}\right)\right]
+ \\  \\
\nonumber
\int_0^{2\theta_1}{d\vartheta_1\,\vartheta_1\over\theta_1^2}
\int_0^{2\theta_2}{d\vartheta_2\,\vartheta_2\over\theta_2^2}
\left[S_+\left({\vartheta_1 \over \theta_1}\right)S_+\left({\vartheta_2 \over \theta_2}\right)
C'_{++}(\vartheta_1,\vartheta_2)\right].
\end{eqnarray}
}}
Eq. (\ref{bigcov}) is  the Eq. (42) in \cite{SCHNEI02} transposed for
the dispersion $\langle \gamma^2\rangle$.
The first term is a pure
Poisson noise term, due to the intrinsic ellipticities of the galaxies.
Its value decreases rapidly with the scale $\theta$. 
The second term is determined by
$C'_{++}(\vartheta_1,\vartheta_2)$, the shear correlations dependent part
of $C_{++}$ (the covariance matrix of $\xi_+$). It is
computed using equations (32) and (34) of \cite{SCHNEI02} with the
the shear correlation function computed from the 
 convergence power spectrum at the fiducial model as,
\begin{equation}
\xi_{\pm}(\theta)={1\over 2\pi} \int_0^\infty{\rm d} k \,
k P_\kappa(k) J_{0,4}(k\theta).  
\end{equation}
We do not include extra sources of errors, in particular 
we assume that galaxy shape measurement and PSF anisotropy corrections 
are free of systematics.
\begin{figure*}
\centering
\includegraphics [height=12cm] {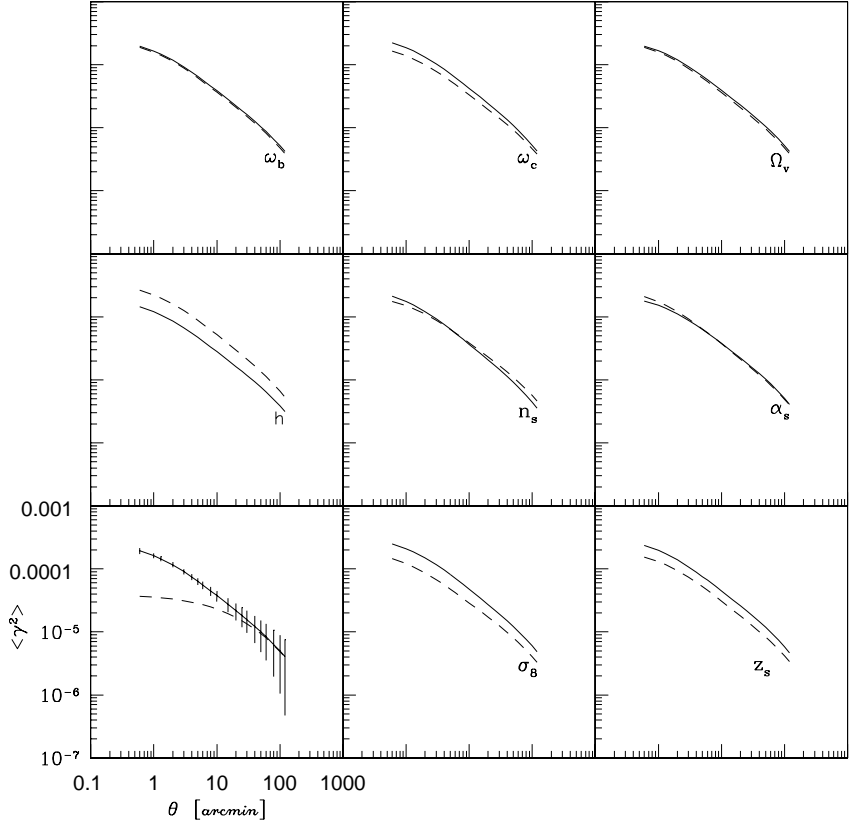}
\caption{Shear variance as a function of scale. The bottom left plot shows
the fiducial model with $3\,\sigma$ error bars. 
In the other 8 plots, we change the labelled parameter 
by a step of $+10\%$ (solid line)
and $-10\%$ (dashed line), except for $\alpha_s$ where the step
is $\pm 100\%$. In each case, the other 7 labelled parameters are kept
constant. Flatness is not imposed.}
\label{var}
\end{figure*}

Table \ref{manip} summarizes the survey properties.
\begin{table}
\caption{Cosmic shear : survey specifications.}
\label{manip}
\begin{center}
\begin{tabular}{ll}
\hline \hline
 Size of the survey: &A=$170\,{\rm deg} ^2$ \\
 Density of galaxies: & $n_g=20\,{\rm arcmin}^{-2}$ \\
 Intrinsic ellipticity dispersion:& $\sigma_\epsilon=0.4$ \\
 Scales probed:& $0.6\,'<\theta<2\deg$\\
 \ & $90<l<18000$\\
 \hline \hline
\end{tabular}
\end{center}
\end{table}
These are based on real observations;
the values for $\sigma_\epsilon$ and $n_g$ (the effective density, once galaxy
selection is done) are found in
cosmic shear surveys (\cite{VW02}) and the field size is the total
size of the CFHT Legacy Survey. To choose the upper limit on the 
angular scale, we notice, from Eq. (\ref{bigcov}),
that the computation of the covariance of the shear dispersion at  
a scale $\theta$ involves an integration up to $2\,\theta$.
Furthermore, the integrations involved in the computation of $C'_{++}$
need an extra factor of $\sqrt{2}$. We define $\theta_{max}$ to be such 
that $2\,\sqrt{2}\,\theta_{max}$ corresponds to the largest wavelength
that can fit in a field of the survey area. Given that the fields of the
CFHTLS-Wide survey have an approximate size of $7\degr \times 7\degr$,
we find $\theta_{max}$ to be around $2\,\degr$.
The lower limit on the scale probes the deep non-linear regime.

The solid line in the bottom left plot of Fig. \ref{var} shows
the shear variance of the fiducial model (Table \ref{fiduc})
as function of angular scale, along with
$3\sigma$ error bars computed from Eq. (\ref{bigcov}) at 20 angular scale points
ranging from $0.6\, {\rm arcmin}$ to $2\degr$. The error bars are smaller
at intermediate scales, slightly larger at the smallest scales, where they are
determined by statistical noise, and are noticeably bigger at the
largest scales, which are cosmic variance dominated.
They are slightly optimistic at small angular scales, due to the 
Gaussianity assumption, made in the derivation of the covariance, lacking
the non-linear enhancement of the signal
The dashed line 
shows the shear variance without the non-linear corrections.
Both lines become well separated below $10 {\rm ~arcmin}$.
The other panels on Fig. \ref{var} illustrate the cosmic shear
sensitivity to different cosmological parameters.
From this figure, it is clear that cosmic shear is more sensitive to
$\Omega_m$,  $\sigma_8$ and the source redshift, with the cosmic shear signal
increasing with the increase of these parameters. It also shows that the 
dependence on $n_s$ is a stronger function of scale than for the other parameters.
This is in agreement with
theoretical expectations derived from linear 
perturbation theory (\cite{BERNVWM97}):
\begin{equation}
\label{varapprox}
\langle \gamma^2(\theta) \rangle^{1/2} \propto \sigma_8 \,\Omega_m^{0.75}
\,z_s^{0.8}\,\left({\theta\over 1'}\right)^{-(n_s+2)/2}.
\end{equation}
The measurement of the signal down to small (non-linear) scales introduces
additional parameter sensitivity 
with some degeneracies broken, like
for instance the parameter pair ($\Omega_m$, $\sigma_8$), as shown
in \cite{JS97}. This gain in sensibility is also noticeable in 
the $\alpha_s$ case.

\section{Markov chain Monte Carlo}
\subsection{General considerations}
The probability distribution function (PDF) of an $m$-dimensional vector parameter 
 $\bf{p}$ given the $n$-dimensional vector data
$\bf{x}(\bf{p}_0)$ (the posterior PDF $P(\bf{p}|\bf{x})$), can
 be calculated using Bayes theorem
from the prior PDF $P(\bf{p})$ and the conditional PDF of the data
  given the parameter vector (the likelihood $L(\bf{x}|\bf{p})$) :
\begin{equation}
\label{bayes}
P(\bf{p}|\bf{x})={{P(\bf{p})L(\bf{x}|\bf{p})}\over{\int{P(\bf{p})L(\bf{x}|\bf{p})d\bf{p}}}}.
\end{equation}

Its analytical computation would involve high-dimensional integrations,
not only to compute the normalisation (also called the evidence or the
marginalised posterior) but also to extract information 
from the posterior, such as
the determination of means or marginalised lower dimensional distributions.
Thus, Eq. (\ref{bayes}) is analytically solved only for special cases:
For Gaussian PDFs, with mean given by $\bf{p}_0$, 
 we see from Eq. (\ref{bayes}) that the posterior inverse covariance matrix 
  is $C^{-1}_{ij}=F_{ij}+F^{prior}_{ij}$, where $F$ is the Fisher information 
  matrix,
\begin{equation}
\label{fishdef}
F_{ij}=\left<{{\partial^2\cal L}\over{\partial p_i \partial p_j}}\right>_{\bf{x}}
\equiv \left({{\partial^2\cal L}\over{\partial p_i \partial p_j}}\right)_{\bf{p}=\bf{p}_0}.
\end{equation}
In general, for non-gaussian PDFs, this Fisher matrix method still gives
valuable information, since
we can always Taylor-expand $\cal L=-\ln$~$L$ around the point 
of maximum likelihood, $\bf{p}=\bf{p}_0$, obtaining, to quadratic order,
\begin{equation}
\label{quadf}
\Delta\cal L(\bf{p})=(\bf{p}-\bf{p}_0)^tF(\bf{p}-\bf{p}_0),
\end{equation}
and use the Fisher matrix as a linear approximation to the inverse covariance matrix.
Being a covariance matrix of a real physical problem, 
$F^{-1}$ must be positive-definite. 
Therefore an hypersurface of constant $\Delta\cal L(\bf{p})$, as defined by 
 Eq. (\ref{quadf}), is an hyperellipse which will be an approximation of a 
 certain $\sigma$ volume of the posterior.
The one-dimensional parameters marginalised errors, i.e., integrated over
the other components of the parameters vector, are given
by $\Delta p_i (1\sigma)=(F^{-1}_{ii})^{1/2}$. From Eq. (\ref{quadf}) 
 it is clear that $\Delta p_i (1\sigma)=(F_{ii})^{-1/2}$ is the error 
   on $p_i$ when all parameters but $p_i$ are fixed at $\bf{p}_0$.
Note that, since the Fisher matrix is exactly the quantity
involved in the Rao-Cram\'{e}r inequality, the errors computed from the 
 Fisher matrix approximation are always lower limits of the true ones.

In practice, in order to obtain a more precise result,
 the problem is usually solved by 
   computing the posterior at optimized sample points that 
   pave the parameter space. 
   The traditional approach uses a regular grid. 
    This is a computer-intensive procedure with computation time
rising exponentially with the space dimension, which limits the number of 
  parameters that can be explored.
 Markov chain Monte Carlo sampling (\cite{MCMC1}) overcomes this limitation.

The use of the MCMC technique in cosmological parameters estimation was first 
  implemented in \cite{CHRIS01}, following the proposal of \cite{CHRISMEYER01}. 
   Current tools like CosmoMC (\cite{LEWISBRI02}) no longer evaluate
   the likelihood 
   at fixed points but at selected positions of a Markov chain. 
 Each chain point, $p_{i+1}$, is derived from the previous chain point,
$p_i$, in such a way that the transition probability from $p_i$ to $p_{i+1}$ 
times the posterior PDF of $p_i$, equals the product of the 
transition probability from $p_{i+1}$ to $p_i$ by the posterior PDF of
$p_{i+1}$. Thus, after a relaxation time, the chain reaches the
equilibrium and constitutes a sample of the posterior.
A clear advantage of this is that statistical properties of the 
 distribution, like the
mean of a parameter or a marginalised confidence interval, can be directly 
  derived from  discreet sample points,  without need to use the
    computed values of the likelihood.
Different priors may also be 
introduced adapting the weighting
scheme defined by the sample, without need to build a new chain.

 The computing time is determined by the number of
points needed to converge to the equilibrium distribution.
If the chain is built in an efficient way, CPU time scales linearly with the 
 dimension of the parameter space. Computing time may be reduced by 
 finding an analytical expression for the posterior. For example, the
 Markov chain data is used to fit the log-likelihood
with a polynomial in \cite{SANDVIK03}.

\subsection{MCMC Analysis}
The MCMC code we developed is  based on the Metropolis-Hastings 
 algorithm (\cite{METROPOLIS} and \cite{HASTINGS}),
like CosmoMC (\cite{LEWISBRI02}), Cog (\cite{SLOZHOB03}) or 
the AnalyzeThis (\cite{DOMU03}) public software.

\subsubsection{Chain progression rules}
We start several chains at different initial positions chosen randomly inside
the limited part of the 7-dimensional parameter space we aim to explore (Table
\ref{cslimit}).

\begin{table}[h]
\caption{Parameters and exploration range investigated by the 
  MCMC chain : The upper part of the table shows the 
  7 parameters used in the initial proposal density, along with their
  exploration range. The bottom part shows extra imposed limits to the
  MCMC exploration.
Other independent cosmological parameters are kept constant at their
fiducial values.}
\label{cslimit}
\begin{center}
\begin{tabular}{llll}
\hline \hline
$\omega_b$ && 0.01 & 0.04\\
$\omega_c$ && 0.01 & 0.3 \\
$h$ &&0.4 & 1.0 \\
$n_s$ &&0.5 &1.4 \\
$\Omega_v$ &&0.3 & 0.9 \\
$\alpha_s$ &&-0.2 & 0.2 \\
$\sigma_8$ &&0.6 & 1.2 \\
\hline
$\Omega_K$ &&-0.1 &0.1\\
$Age$ &(Gyr)&10 &20\\
$\Omega_c$&$<$&3$\Omega_b$&\\
\hline \hline
\end{tabular}
\end{center}
\end{table}

The next  point is proposed using a proposal PDF, $q(p_{i+1}|p_i)$ and the 
 unnormalised posteriors of
both points are compared to decide if the new point is acceptable.
The PDF acceptance rule we use for the next step point 
   $p_{i+1}$ is defined by   
\begin{equation}
\label{accept}
\alpha(p_{i+1}|p_i)=min\left\{
{
{p(p_{i+1})L(x|p_{i+1})q(p_{i}|p_{i+1})}
\over{p(p_{i})L(x|p_{i})q(p_{i+1}|p_{i})}
},1\right\}.
\end{equation}
 $\alpha$ is compared to a random number $u$ generated from a 
 uniform distribution  between 0 and 1. If 
  $\alpha>u$, then the  new $p_{i+1}$ is accepted; otherwise
  it is rejected and we keep the point $p_{i+1}=p_i$.
 Since both the proposal and the acceptance densities only depend on the 
  current element of the chain and not on its previous history,  the 
   resulting sample will be a Markov chain.
It is worth noticing that the 
  ratio in this definition makes the normalisation of the posterior 
 PDF unnecessary. Furthermore, the presence of the proposal PDF in 
 Eq. (\ref{accept}), ensures 
  that $p(p)L(x|p)$ is the equilibrium
function, even in the case of a non-symmetrical $q$, {\em i.e.} when the 
  probability of proposing $p$
from $p'$ is different from that of proposing $p'$ from $p$.

The result is independent of the proposal density, $q$.
We use as $q$, at the beginning of the chain, 
a 2-dimensional Gaussian distribution centered at the current chain element.
 Hence, only 2 of the 7 parameters (randomly chosen) change at each step. 
The covariance matrix of $q$ is chosen to be
of the order of the expected, squared, $1\sigma$ error bars.
The $1\sigma$ error 
value is used as the step definition criterion and guarantees the 
  step amplitude has an adequate size. Would it be too small,  
 the chain would move too slowly and could never leave the vicinity of the
 best fit.  This situation is known as poor mixing and leads 
to underestimated confidence limits.
 In contrast, if the proposed steps are too large, the acceptance rate
will be too small and once again the chain will move slowly.

In order to have an adequate initial proposal density we derived approximate
$1\sigma$ errors from a Fisher matrix computation. Applying Eq. (\ref{fishdef})
to Eq. (\ref{like}) leads to (\cite{TEGTAHE97}),
\begin{equation}
\label{fisher}
F_{ij}=\sum_{\alpha=1}^n\,\sum_{\beta=1}^n C^{-1}_{\alpha\beta}\,
\left(
{{\partial\langle \gamma^2\rangle_\alpha}\over{\partial p_i}}
{{\partial\langle \gamma^2\rangle_\beta}\over{\partial p_j}}
\right),
\end{equation}
where the derivatives of the shear variance are evaluated at the fiducial
model. To numerically evaluate these derivatives, we compute the shear variances
at points $p_i^{fid}\pm\,\Delta\,p_i$, with the deltas ranging from
 $0.1\%$ to $10\%$ depending 
 on the sensitivity of the shear statistic to each cosmological parameter, $p_i$. 
  However, computing and sampling errors, coming mainly from the near cancellation
  of certain combinations of derivatives, as was pointed out by \cite{EIHUTEGetal98},
   generate small fluctuations in the Fisher matrix coefficients 
   that are amplified by matrix inversion when the covariance matrix 
    is derived from the Fisher matrix. 
   For this reason, we did not use the  inverse Fisher matrix as 
   the covariance matrix of the initial proposal density,
  but used instead a diagonal covariance matrix whose ratios between coefficients are equal to the ratios
  between the diagonal coefficients of the inverse Fisher matrix.
It is worth noticing that
one way to obtain a not so nearly singular Fisher matrix is to include the
derivatives of the data covariance matrix in the derivation of the Fisher
matrix formula (Eq. (\ref{fisher})).    

In order to better explore the directions of degeneracy 
and consequently speed up the convergence, 
a non-diagonal proposal covariance matrix is needed. Hence, 
after $1000$ steps, the covariance matrix of the chain in progress
is computed. From it, a new set of 7 parameters, aligned with
the eigenvectors of the evaluated correlation matrix, is defined.  
From that step on, the new sample points are built from
a combination of 2 eigenvector directions, randomly chosen at each step.
 Though it defines the next direction, the step size does not
necessarily needs to match the corresponding eigenvalues.
In fact, after 1000 steps the covariance matrix is smaller than it will be
at its converged value, so we must scale it.
  We update periodically the proposal covariance matrix.
  However, since this process computes a new sample covariance matrix, 
  it cannot be done too frequently; otherwise the progression of the 
   chain too much depends on the previous elements and would no longer 
    be a Markov chain. After a few periods we freeze the proposal density
    and set the multiplicative correction factor 
 to 1. The optimal scaling value depends on the number
 of dimensions probed (\cite{MCMC8}).

\subsubsection{Convergence and goodness}
The first elements of a chain depend on the starting point. It is only 
 after a so called burn-in period that the chain starts sampling the 
  target density. 
 Although it is not possible to say with certainty that a finite sample 
  from an MCMC algorithm is representative of the target distribution, 
   several tests have been proposed to detect non-convergence of the chain 
    (for a review see \cite{COWCAR94}).  In this work, 
     we used the Gelman and Rubin test to estimate the size of the burn-in.

Let us consider a Markov chain of a given parameter composed of $2n$ iterations.
Due to the random selection process of the starting point we
expect $m$ different chains to differ significantly at the beginning and 
to converge towards the same distribution as the number of iteration steps
increases. The burn-in interval is set when the typical separation of
several chain points at a given iteration is similar to the amplitude
of the chain internal fluctuations.
 At each iteration $i$, two quantities can be computed for each parameter:
\begin{figure}[t]
\centering
\includegraphics [height=8cm] {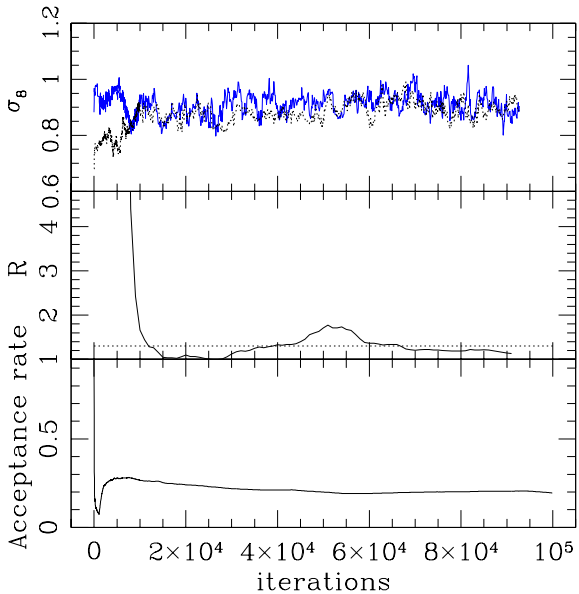} 
\caption{Monitoring a chain.The upper plot shows the successive values of $\sigma_8$ for two 
of the CFHTLS simulated chains with different starting points.
In the middle plot, the Gelman and Rubin test between these 2
chains is made at several stages. This example shows that it is possible to detect an apparent
convergence around the iteration step $20 \times 10^3$.
From the iteration $60 \times 10^3$ on, $R$ started to consistently approach 1
for all  parameters, leading us to choose a burn-in of $30 \times 10^3$.
The horizontal line is $\mbox{R=1.3}$.
In the bottom plot, the acceptance rate of one chain is computed at each iteration $i$,
using all chain elements from the first one up to $i$.
In the very beginning we notice two opposite behaviors : a shear drop and a
steady raise. The transition takes place at the moment the proposal density starts to follow the
covariance matrix of the sample, showing the efficiency of that procedure. Afterwards, the
acceptance rate drops slowly stabilizing at about $20\%$.
}
\label{monitor}
\end{figure}
\begin{itemize} 
\item The within-chain variance, $W(p)$, is the average of the $m$ variances of the parameter,
\begin{equation}
W(p)={{1\over m}\sum_{j=1}^m\,{1\over{n-1}}\sum_{i=n}^{2n}\,
(p_i(j)-<p>_i(j))^2}.
\end{equation}
\item The between-chains variance, $B(p)$, is the variance of the parameter means from the $m$ chains,
\begin{equation}
B(p)={{1\over{m-1}}\sum_{j=1}^m\,(<p>_i(j)-<<p>_i(j)>_j)^2}.
\end{equation}
\end{itemize}
In the first iterations, each chain is concentrated in a different starting 
 region, so $B$ is in most cases much larger than $W$. When the iterations 
  increase, $W$ grows while $B$ decreases and get closer and closer 
 to zero at convergence. Since it is not possible to explore  
 the whole space, $W$ is always an underestimate of the within-chain variance.
 Hence, $B+W$ is an overestimate of the parameter variance.
In the limit $i\rightarrow \infty$, both estimators approach the true variance 
 from opposite sides. Their ratio,
\begin{equation}
R={1\over W}\left(B+{W\,n\over{n-1}}\right),
\end{equation}
is therefore a suitable estimate to monitor the convergence.

When, after $2n$ iterations, $R$ is close to one for all the
quantities of interest, {\em i.e.},
for all the parameters and derived parameters we
want to analyse, we assume the chain has converged.
The first $n$ iterations are the burn-in period. They are
discarded and the actual
marginalised posterior density of a parameter is drawn by its 
frequency of appearance in each bin, during the iterations
$n+1$ to $2n$.
When after $2n$ iterations
there was not enough time to explore the tails of the distribution
the errors of the target distribution are underestimated.
Therefore, it is useful to let the chains run for a longer period in 
order to
 get a better mixing. In the process, the value of the
estimate $R$ may raise before getting smaller again. We show an example
of this situation in Fig. \ref {monitor}, where we
follow a chain evolution. 

Once a chain has stopped, we must remove the residual
correlation between the consecutive elements.
For this reason it is recommended to thin the chain out, {\em i.e.},
to keep only 1 out of $k$ consecutive elements.
The most widespread method in the literature to determine the
thinning factor, $k$, is the Raftery and Lewis method (RL)(\cite{MCMC7}).
\begin{figure}[h]
\centering
\includegraphics [height=7cm] {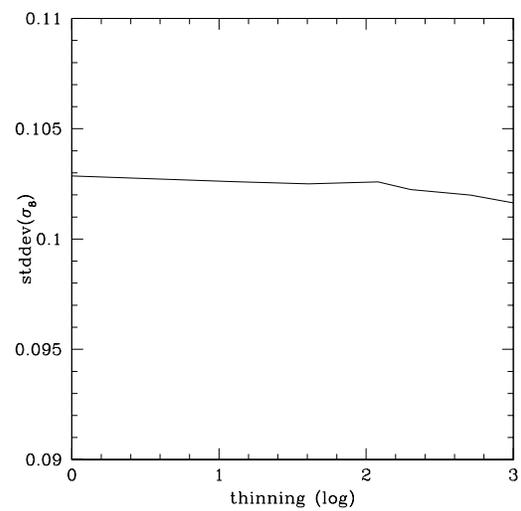} 
\caption{The standard deviation of $\sigma_8$ computed from different samples taken from the same
chain by using different thinning factors is plotted against the thinning in logarithmic
scale.
}
\label{thin}
\end{figure}
This method starts by constructing several chains from the converged MCMC chain,
by thinning the latter with several different values. 
A weight may be assigned to each one of 
the thinned chains, according to its compatibility to an independence
chain (a chain with no correlation between its consecutive elements).
RL computes the weight of a chain from
the ratio between its evidence and the evidence of an independence
chain. The evidences are computed in the Bayesian Information
Criterium (BIC) form of \cite{SCHWARZ78}, which is a gaussian approximation 
that may be derived from the Bayes formula (Eq. (\ref{bayes})). It writes,
$\rm{ratio}=\rm{BIC}=G^2-2\,{\sc log}\,n$, where $n$
is the number of elements in the chain. The $G^2$ statistic is a 
$\chi^2$ that measures the fit of a chain to an independence chain.
To obtain $G^2$, one counts the number of transitions
between bins of the chain, in order to get the ratio between the probability
of the chain to have a certain value at a certain step $i$ for a given 
chain value of a previous step $i-j$, and the probability independently
of the value at $i-j$.
In practice, when counting the transitions, only 2 bins are assumed,
{\em i.e}, a chain element becomes a 0 or a 1 as whether its parameters
values are
less or greater than a certain cut-off, that we choose to be the parameters
$2\sigma$ values. The greatest weight is attributed to the longest chain
verifying $BIC<0$. Its thinning value is the obtained $k$ factor and 
that chain is the best-fit to an independence chain that can be obtained by
thinning the original chain.

It is worth noting that this is an application of the use of the evidence
to the problem of selection between different sets. Another evidence based
weighting scheme is used when combining different sets of  
cosmological data in \cite{HOBSONetal02}.

There are other methods
to estimate a thinning factor.
In particular, \cite{TEGSDSS03} obtain it by
defining a more intuitive chain correlation length. 
Using RL, we  
obtained very large thinning factors for some of the MCMC chains (of order 100).
However, we checked the dependence of the results on the thinning factor, by
computing the parameters confidence levels using several chains, thinned
from the original chain with different values of $k$ and
found no appreciable difference in the results
(see Fig. \ref{thin} for an example).
Hence, in order not to lose so many chain elements we did not use
the Raftery and Lewis method results, but a simpler criterium: for each chain
we choose the thinning factor as the average
multiplicity of the chain elements.

The results in the next section are computed from 4 chains with $10^5$
elements each, with a burn-in size of $30\times 10^3$ and a thinning
factor of 5. The chains were merged, leading to a final sample with
about 50\,000 elements.

\section{Cosmological parameters from CFHTLS cosmic shear statistics only}
\begin{table*}
\caption{Numerical results for the cosmic shear sample, including $1\sigma$ precision on
individual parameters and a principal components analysis. 
I: $68\%$ confidence levels
in absolute value (first line) and in percentage of the corresponding
parameter fiducial value (second line).
II: The 7 eigenvectors of the correlation matrix ordered by decreasing accuracy.
The column named $1\sigma$, lists the dispersion of each $X$  parameter defined
in Eq. (\ref{eigen}), which is equivalent to the square root of the
corresponding eigenvalue. Each line $i$ shows the coefficients $a_{ij}$ of
Eq. (\ref{eigen}), {\em i.e.}, the projections of the corresponding
$X_i$ on each of the 7 parameters $p_j$ labeled on the very top of the table.
Naturally the derived parameter $\Omega_m$ is not used for the computation of the
eigendirections and $\sum_{j=1}^7\,a_{ij}^2=1$.
III: The 7 eigenvectors computed for mean subtracted data normalized by the means.
The parameters used are $(\Omega_b,\Omega_m,h,n_s,\Omega_v,\alpha_s,\sigma_8)$.
The column $1\sigma$, lists the dispersion of each $Y$ parameter defined
in Eq. (\ref{eigeny}). The next 7 columns show the components, $b_{ij}$, of the $Y_i$.
In the last column we show the relative contribution of the main parameter
involved in each pc.
IV: Each line shows the fractional error of each of the 7 parameters
$(\Omega_b,\Omega_m,h,n_s,\Omega_v,\alpha_s,\sigma_8)$ computed 
using a limited number of principal components. The first line, 
$\#pc=1$, refers to using only the projections of $Y_7$. In $\#pc=2$ both $Y_7$
and $Y_6$ are used. Using all the PCS, we recover, in the last line, $100\%$
of the error values for all parameters.
}
\label{tabeigen}
\begin{center}
\begin{tabular}{ccccccccccc}
\hline \hline
I & & & $\omega_b$ & $\omega_c$ & $h$ & $n_s$ & $\Omega_v$
& $\alpha_s$ & $\sigma_8$ & $\Omega_m$\\
\hline 
& & & 0.020 & 0.047 & 0.129 & 0.176 & 0.155 & 0.073 & 0.104 & 0.067\\
& & & 93.1 & 41.3 & 18.2 & 18.9 & 21.3 & 182.5 & 11.6 & 24.8 \\
\hline \hline
II & pc & $1\sigma$\\
\hline 
& $X_1$ & 0.061 & 0.218 & 0.616 & -0.665 & -0.102 & 0.156 & -0.023 & 0.308 \\
& $X_2$ & 0.104 & -0.289 & 0.426 & 0.153 & 0.792 & -0.149 & 0.246 & 0.042  \\
& $X_3$ & 0.368 & 0.412 & -0.341 & -0.099 & 0.068 & -0.314 & 0.661 & 0.406 \\
& $X_4$ & 0.805 & -0.370 & -0.008 & -0.040 & -0.160 & -0.693 & -0.376 & 0.462  \\
& $X_5$ & 1.142 & -0.259 & 0.066 & 0.425 & -0.188 & 0.522 & 0.128 & 0.651 \\
& $X_6$ & 1.271 & 0.703 & 0.206 & 0.465 & 0.161 & -0.100 & -0.428 & 0.166 \\
& $X_7$ & 1.811 & 0.004 & -0.525 & -0.354 & 0.520 & 0.304 & -0.402 & 0.271 \\
\hline \hline
III & pc &  & $\Omega_b$ & $\Omega_m$ &  &  & &  &  & $\% mp$ \\
\hline 
& $Y_1$ & 0.007 & 0.009 & 0.404 & -0.040 & -0.070 & 0.207 & 0.000 & 0.887 & 78\\
& $Y_2$ & 0.022 & -0.082 & 0.167 & 0.388 & 0.889 & -0.143 & 0.027 & 0.046 & 79 \\
& $Y_3$ & 0.103 & 0.084 & -0.490 & -0.287 & 0.088 & -0.714 & 0.068 & 0.384 & 51 \\
& $Y_4$ & 0.193 & 0.124 & -0.231 & 0.872 & -0.357 & 0.141 & 0.003 & 0.148 & 76 \\
& $Y_5$ & 0.302 & 0.175 & -0.677 & -0.031 & 0.246 & 0.634 & 0.124 & 0.176 & 46\\
& $Y_6$ & 0.811 & -0.904 & -0.235 & 0.041 & -0.044 & 0.055 & -0.332 & 0.102 & 82\\
& $Y_7$ & 1.936 & 0.350 & -0.038 & -0.025 & 0.080 & 0.007 & -0.932 & 0.176 & 87\\
\hline \hline
IV & \# pc \\
\hline
& 1 && 67.7 & 24.7 & 27.1 & 81.5 & 6.9  & 98.9 & 29.6 & \\
& 2 && 99.8 & 68.7 & 32.7 & 83.6 & 22.1 & 99.9 & 77.5 & \\
& 3 && 99.9 & 97.3 & 33.1 & 92.4 & 92.8 & 100  & 90.2 & \\
& 4 && 99.9 & 98.5 & 98.5 & 99.3 & 93.7 & 100  & 93.5 & \\
& 5 && 100  & 99.9 & 99.9 & 99.4 & 99.9 & 100  & 99.6 & \\
& 6 && 100  & 99.9 & 99.9 & 99.9 & 99.9 & 100  & 99.7 & \\
& 7 && 100  & 100  & 100  & 100  & 100  & 100  & 100  & \\
\hline \hline
\end{tabular}
\end{center}
\end{table*}

We will now extract information about the cosmological parameters
from the obtained sample of the posterior PDF.

We start by computing
one-dimensional confidence levels for
the parameters. Table \ref{tabeigen} (I) shows the standard deviations
obtained for a set of 8 parameters;
the 7 original ones and $\Omega_m=(\omega_b+\omega_c)/h^2$.
These values are computed
using all the sample points (hence being marginalised values) and
are shown in absolute value in the first line 
of Table \ref{tabeigen}(I) and in percentage of the parameters fiducial
values on the second line.
Since MCMC probes a non-gaussian posterior, asymmetric error intervals may 
also be computed. We found the positive and negative $68\%$ confidence levels
do not differ much from the standard deviations and do not show them here.
As compared to the early {\sc Virmos}-Descart results, the CFHTLS configuration 
does not seem to increase the precision on $\sigma_8$.
This is however misleading since \cite{VW02} carried out their 
maximum likelihood analysis using only 4
parameters $(\sigma_8\, ,\Omega_m\, ,\Gamma=\Omega_m\,h\, ,z_s)$. 
Hence the actual improvement is eventually much better.

One-dimensional confidence levels do not show the 
detailed statistical structure of the cosmological parameter space.
In the following we describe the interest in 
using a principal components analysis in the cosmological parameter space,
a technique pioneered in \cite{EFSTBOND99}.

\subsection{Principal components analysis}

\begin{figure}[h]
\centering
\includegraphics [height=6cm] {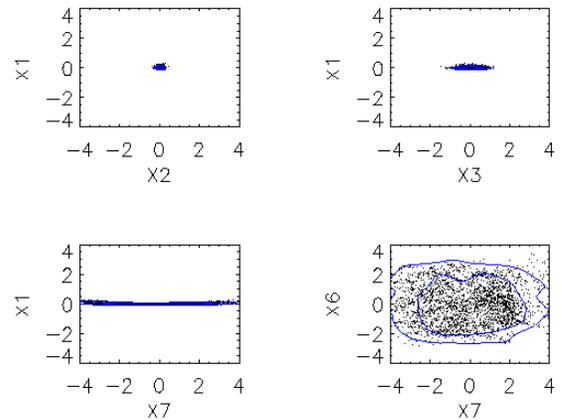}
\caption{1 and $2\sigma$ contours for pairs of principal components. Since they are linear
combinations of standardized variables, they are all centered on zero. It is obvious these
are plots of non-correlated parameters. The low scattering in the $X_1$ direction indicates
that the cosmological parameters combination defined by $X_1$ is strongly constrained.
On the other hand, the high scattering in the principal components associated with the
highest eigenvalues, indicates these components contain most of the sample dispersion,
and determine the cosmological parameters errors.
}
\label{pca2d}
\end{figure}

The principal components of the sample, $X_i$, are derived from the 
eigenvectors of the sample correlation matrix. The correlation matrix is the
covariance matrix of the sample of parameters in standardized form, which means
each parameter value is rescaled by subtracting the mean and dividing by the 
dispersion. The principal components (PCS) can 
be expressed as a linear combination of the 7 rescaled parameters as follows:
\begin{equation}
\label{eigen}
X_i=\sum_{j=1}^7\,a_{ij}{(p_j-<p_j>)\over \sigma_{pj}},
\end{equation}
where $<p_j>$ and $\sigma_{pj}$ are the mean and the dispersion of the 
parameter $p_j$.The coefficients $a_{ij}$  are listed in Table \ref{tabeigen} (II), where
the principal components are ordered by decreasing accuracy. Figure \ref{pca2d}
shows some examples of 2-dimensional plots of the parameters
defined by Eq. (\ref{eigen}).
\begin{figure*}
\centering
\includegraphics [height=13cm] {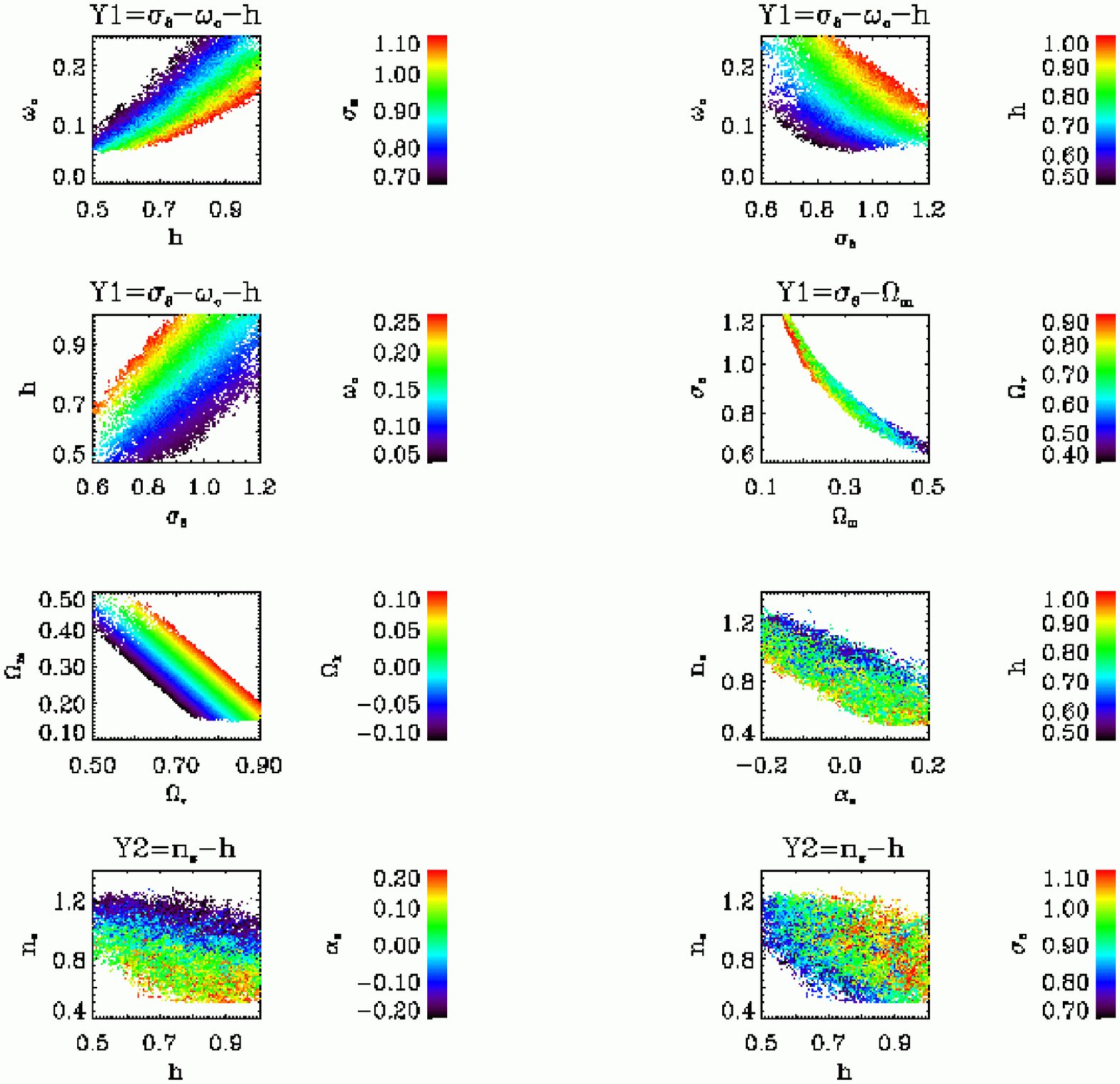} 
\caption{2-dimensional scatter plots colored by a third parameter putting
in evidence the multi-dimensionality of the degeneracies.
Panels are numbered from top left to bottom right.
Panels 1-3: Illustrate the best determined pc in the form 
$\omega_c-\sigma_8-h$. The well defined continuous gradients in all 3 cases
show the existence of a degeneracy, even though that would not be obvious
from the 2-dimensional projections alone.
Panel 4: Using $\Omega_m$ instead of $\omega_c$, the best determined pc already
defines a narrow degeneracy in the 2-dimensional plane $(\Omega_m,\sigma_8)$.
$\Omega_v$ values are also shown.
Panel 5: Coloring the $(\Omega_v,\Omega_m)$ plot with the curvature values,
a linear pattern appears.
Panel 6: Shows the degeneracy $\alpha_s-n_s$ which was not clear from the 
eigenvectors table alone.
By coloring with $h$ it also shows that this degeneracy contributes to $Y_2$.
Panels 7,8: The second best determined pc relates mainly $n_s$ and $h$. 
These plots show it also has some correlation with $\alpha_s$ (panel 7)
and less from $\sigma_8$ (panel 8).
} 
\label{color}
\end{figure*}

The most accurate PCS are the best determined quantities by the 
CFHTLS-wide cosmic shear experiment. In order to see to which combinations 
of cosmological parameters they correspond, 
one can look at the eigenvectors components, {\em i.e.}, a high coefficient
$a_{ij}$ means $p_j$ strongly contributes to $X_i$. 
However, since we are working with standardized parameters,
the direct
reading of the coefficients may be misleading. In fact, if we take any subset
of 2 parameters and compute its eigenvectors, they both will have equal 
components, which obviously does not mean each principal component
has equal contributions
from each parameter. Hence, for the
purpose of obtaining a set of 
meaningful coefficients
, it is adequate to rescale the parameters
differently. We rescale them 
by subtracting the mean and dividing by the mean. This is refered to as the
fractional data in \cite{CHU03}.
The covariance matrix of the
rescaled sample relates to the original covariance matrix as,
$C'_{ij}=C_{ij}/(<p_i><p_j>).$

We compute a set of principal components, $Y$, from the fractional covariance matrix,
expressed as,
\begin{equation}
\label{eigeny}
Y_i=\sum_{j=1}^7\,b_{ij}{(p_j-<p_j>)\over <p_j>}.
\end{equation}
Each $Y_i$ approximately corresponds to one $X_i$ but they are not equivalent since
the principal components depend on the scaling of the variables. We show this
set in Table \ref{tabeigen} (III), computed for a slightly different 
set of cosmological parameters with 
$(\omega_b,\omega_c)$ replaced by $(\Omega_b,\Omega_m)$.

\subsubsection{Well constrained PCS}

The components of an eigenvector explicitly show
the contribution of each parameter to a principal component:
\begin{itemize}
\item The first one, $Y_1$, is dominated by  contributions from 
 $\sigma_8$ and $\Omega_m$. This is the well known $\sigma_8-\Omega_m$ 
  degeneracy. $\Omega_v$ is  also  non-negligible. Its presence here comes 
  from its correlation with $\Omega_m$ 
 through the prior on the curvature used to compute the chains 
  (see Table \ref{cslimit}).
\item The second best constrained direction, $Y_2$, couples the primordial 
 spectrum index, $n_s$, with $h$ and $\Omega_m$. This comes from the correlation 
 between the tilt $n_s$ and the parameterisation of the slope of the mass
 power spectrum to which the
 shear is sensitive, $\Gamma=\Omega_m \,h$. 
Notice the coefficients of $n_s$ and $h$ have the same sign,
as is also the case with $\sigma_8$ and $\Omega_m$ in $Y_1$, which shows the
orientation of the degeneracy. 
\end{itemize}
 The relative contribution of the main parameter involved in each principal component
  is in the last column of Table \ref{tabeigen} (III).
 For example, the projection of $Y_1$ on the direction of its main
parameter ($\sigma_8$) is $78\%$. Although these fractions may be important,
 they are significantly below 100\%, showing that none of these 
  principal components that describe the cosmic shear spectrum 
   depends on  a single cosmological parameter.

The 2-dimensional projections of Fig. \ref{color} use 
colors to produce 3-dimensional  plots that better describe sensitivity
 of cosmic shear to cosmological parameters. The comparison 
 of Table \ref{tabeigen} (II) and Table \ref{tabeigen} (III) 
 shows for example  that 
 the $\sigma_8-\Omega_m$ degeneracy is equivalent to a
$\omega_c-h-\sigma_8$ degeneracy. A scatter colored plot, like 
  the top panels of  Fig. \ref{color}, illustrates this in a simple way.
 On this figure, we plot the sample points for the 3 possible pairs of 
  parameters, colored by the one left out. The continuous gradient along the 
   third parameter is obvious.  In particular, the degeneracy that is hidden in 
    the $\sigma_8-\omega_c$ plane becomes evident once
the points are colored according to $h$. Likewise, 
 the cosmic shear $\sigma_8-\Omega_m$ degeneracy  pattern 
  $\sigma_8 \propto \Omega_m^{-0.5}$ is shown on the fourth panel,
  with a continuous gradient along $\Omega_v$. The fifth panel shows how this
  correlation between $\Omega_m$ and $\Omega_v$ is related to the curvature. 

The color plots also help in understanding degeneracies derived from 
 the analysis of Table \ref{tabeigen}.  The bottom panels of 
    Fig. \ref{color}, illustrate the cases on 
     the $Y2$ term either when a third parameter does not contribute to the degeneracy 
 ($\sigma_8$) (producing a mixture of colors), or when it does 
 ($\alpha_s$). The analysis of Table \ref{tabeigen} alone is indeed confusing:
  while from \ref{tabeigen} (III) it is
not evident that $\alpha_s$ contributes more to the second principal component
than $\sigma_8$ does,  Table \ref{tabeigen} (II) shows what really happens. 
 The need for a careful interpretation of this table is of primarily 
 importance when  fractional
 covariance matrix is used for parameters with fiducial values close to zero.
The simple extraction 
of  eigenvectors components to find degeneracies only provides qualitative
insight since there is no unique set of principal components. 
  For these ambiguous cases, color plots are very useful. The  bottom 
   right panels of Fig. \ref{color}, reveal the sensitivity of $Y2$ to 
   $\alpha_s$ in a much better way than the tables do.

From the MCMC and the principal components analysis, it is 
 possible to describe some cosmic shear denegeracies with empirical laws, 
  using the best determined components $Y_1$ and $Y_2$.  
 Since the 2-dimensional contours of $Y_1$ and $Y_2$ are not 
 ellipses,  we made a new eigenvector calculation, using
   the logarithm of the parameters. The laws are established
 for two parameters only, marginalising over the others. 
   This way we found the shapes
in the $\sigma_8-\Omega_m$ and the $n_s-h$ planes to be:
\begin{equation}
\sigma_8\,\Omega_m^{0.52}=0.467 \pm 0.008
\end{equation}
and $h\,n_s^{0.57}=0.67 \pm 0.10$. Defining $\Gamma=\Omega_m\,h$, 
we find a better constrained, $Y_2$ related, two parameter relation, 
\begin{equation}
\Gamma\,n_s^{0.6} = 0.187 \pm 0.037.
\end{equation}

\subsubsection{Poorly constrained PCS}

As for the other principal components:
\begin{itemize} 
\item $Y_3$ is dominated by $\Omega_v$, $\Omega_m$ and $\sigma_8$. 
This constrains the curvature, mixed with an $Y_1$ orthogonal
contribution of $\sigma_8-\Omega_m$, as the coefficients of these 2 parameters
have now opposite signs.
\item 
The projection of $Y_4$ on the $n_s-h$ plane constrains an
orthogonal direction to the $Y_2$ projection.
\item 
$Y_5$ mixes almost all the parameters, while being dominated by an
$\Omega_m-\Omega_v$ plane orthogonal direction to the one defining the
curvature. As we will see next, $Y_5$ alone is almost enough to 
determine the parameters precision.
\item 
The two worst constrained principal directions,$Y_6$ and $Y_7$, 
are the most aligned ones with single parameters, being
strongly dominated by $\Omega_b$
and $\alpha_s$, respectively.
\end{itemize}

Even though the best determined PCS give strong constraints on certain
combinations of parameters, constraints on individual cosmological parameters 
are strongly dependent on the worst determined PCS, exception made for a
cosmological parameter aligned with a well constrained principal component
(from the last
column of Table \ref{tabeigen} (III) we see there is no such case).
In fact, the worst constrained PCS determine the size of the hyper-volume
defined by the sample in parameter space and account for most of the dispersion 
of the sample. This is the reason why $\Omega_b$ and $\alpha_s$, which dominate
$Y_6$ and $Y_7$, are the worst determined parameters (Table \ref{tabeigen} (I)).
To investigate the contribution of each principal component
to the one-dimensional errors on
parameters, we compute the $1\sigma$ marginalised parameters dispersion using
only a restricted number, $n$, of principal components, as:
\begin{equation}
\label{sigfrac}
\sigma^2(p_j,n)=<p^2_j>\sum_{l=7}^{8-n}\,(Y_l\,b_{lj})^2.
\end{equation}
Notice the case $n=1$ uses only $Y_7$. Notice also the need to 
multiply by the parameters means, since we are using
fractional data. Table \ref{tabeigen} (IV) shows the $1\sigma$ marginalised
errors for each parameter as a percentage of the total $1\sigma$ marginalised
errors (shown in Table \ref{tabeigen}(I)), computed in this way, where \#pc=i 
refers to $n=i$ of Eq. (\ref{sigfrac}).
We see that \mbox{$Y_7+Y_6+Y_5$} (\#pc=3) accounts for over 95\% of several
parameters $1\sigma$ values and from \mbox{$Y_7+Y_6+Y_5+Y_4+Y_3$}
all the correct values are obtained.
Hence, we may conclude that
the statistical information of the sample can be reduced to 5 dimensions plus 2 narrow
priors defined by $Y_1$ and $Y_2$. This way, a 5 dimensional MCMC would be enough
to obtain an equivalent sample in a faster way. 

\bigskip{}

\bigskip{}

Finally, the cosmic shear dependence on the
cosmological parameters may be summarized in a single expression analogous to 
 Eq. (\ref{varapprox}).  It is derived by using 
the surface of least dispersion defined on the 7-dimensions surface. It must 
 be orthogonal to $Y_1$, {\em i.e.}, defined by $Y_1=c^{te}$. This 
  constant is proportional to
 the cosmic shear variance  signal from all scales
 (since they were all integrated to compute
the likelihood, it does not explicitly depend on angular scale). 
 We found that
\begin{equation}
\gamma \propto \sigma_8\,\Omega_m^{0.57}\,\Omega_b^{0.007}\,\Omega_v^{0.18}\,
h^{-0.02}\,n_s^{-0.02}\,\alpha_s^{-0.002}.
\end{equation}

The MCMC and principal component analysis of CFHTLS cosmic shear data alone
  can be generalised with joint data sets. As it has been shown in this 
 section, the method provides useful information on degeneracies and 
  principal components and allows a description of orthogonal 
 directions. Used jointly with CMB data sets, we can therefore 
  predict how CFHTLS cosmic shear and CMB data can be used in an 
  optimal way to shrink the exclusion diagrams attached to each 
  cosmological parameters.

\section{Constraints from Cosmic Shear and CMB}
 
We produced a different set of chains, computing the joint likelihood
of the models with respect to the same cosmic shear fiducial data and
CMB data. 
 
We used the WMAP first year data\footnote{http://lambda.gsfc.nasa.gov}: 
the combined TT power spectrum (\cite{HINWMAP03})
and the TE power spectrum (\cite{KOGWMAP03}). Model's likelihoods with respect
to WMAP data were computed
using the WMAP likelihood code
(\cite{VERDEWMAP03}). In order to have
information from smaller scales, we included CBI data\footnote{
http://www.astro.caltech.edu/\~{}tjp/CBI/}: the mosaic odd binning
(\cite{PEARSONCBI03}).
\begin{table}
\caption{One-dimensional results from the joint sample.
The errors are $1\sigma$. Above the horizontal line are the 7
explicitly changed MCMC parameters, while results for some
other popular parameters are shown under the line.
The column
labeled $g_1$ shows the gain in the parameters precision in relation to
the values obtained with the CMB chains.
In the last column, the gain $g_2$ is computed in relation to available
CMB results (taken from Table 8 of \cite{SPERGEL03}, for the case WMAPext). 
}
\label{jointtab}
\begin{center}
\begin{tabular}{lllll}
\hline \hline
& joint& $g_1$ & $g_2$  \\ \\
\hline 
$\omega_b$ & $0.0224 \pm 0.0008$ &1.4 &1.2\\
$\omega_c$ &$0.112 \pm 0.003$ &3.6 \\
$h$ &$0.71 \pm 0.02$ &1.9 &2.9 \\
$n_s$ &$0.94 \pm 0.03$ &1.7 &2.0\\
$\alpha_s$ &$-0.050 \pm 0.018$ &1.7&2.1\\
$10^9\,A_s$ &$2.8 \pm 0.2$ &1.1\\
$\tau$ &$0.24 \pm 0.04$ &1.0&1.7\\
\hline
$\Omega_m$ &$0.26 \pm 0.02$ &2.8\\
$\sigma_8$ &$0.91 \pm 0.03$ &2.5\\
$\Omega_m\,h$ &$0.188 \pm 0.008$ &3.2\\
$\sigma_8\,\Omega_m^{0.5}$ & $0.47 \pm 0.01$ &6.7\,\,(1.1) \\
$\sigma_8\,e^{-2\,\tau}$ &$0.56 \pm 0.03$ &1.8 \\
\hline \hline
\end{tabular}
\end{center}
\end{table}
\begin{figure}
\centering
\includegraphics [height=8cm] {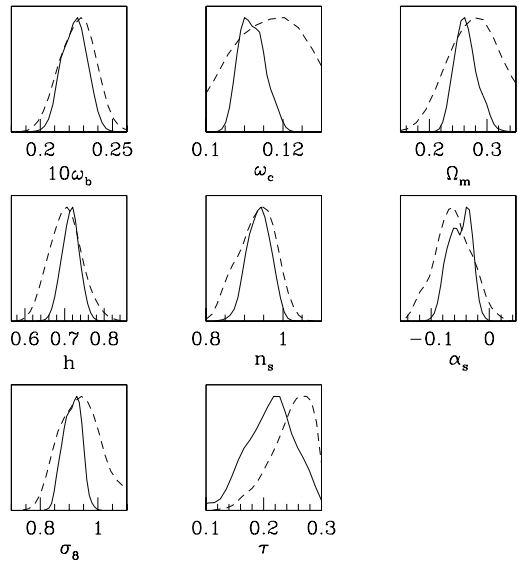} 
\caption{Marginalised distributions for the cosmological parameters
from the joint likelihood chains (solid line) and the CMB chains
(dashed line).} 
\label{cmblens1d}
\end{figure}
The number of independent parameters explored by the
Markov chains was kept at 7: 
$(\omega_b\,,\omega_c\,,h\,,n_s\,,\alpha_s\,,\tau\,,A_s)$.
The normalization is now parameterized by $A_s$, with a
fiducial value of $A_s=2.6\times 10^{-9}$, corresponding in
our fiducial model to $\sigma_8=0.9$.
We restrict to flat models, hence $\Omega_v$ is no longer an
independent parameter and let $\tau$, the
optical depth to reionization, change in the restricted
region of $\tau<0.3$., keeping its fiducial
value at $\tau=0.17$.

We present results from a combination of 8 converged chains about 
$70 \times 10^3$
elements long from which we rejected the first $30\times 10^3$ elements.
The thinning factor is 8, leaving us with a final merged sample of
about 40\,000 elements.

Table \ref{jointtab} shows one-dimensional marginalised results from the joint
sample. In order to explicitly see what can be gained when joining
cosmic shear data to CMB data, we also produced
 CMB only chains. One-dimensional
distributions from both CMB and joint samples are shown in
Fig. \ref{cmblens1d}.
 The ratio between 
the parameters standard deviations obtained with the CMB sample and the 
joint sample, $g=\sigma(CMB)/\sigma(CFHTLS+CMB)$, tells us
to which parameters the combined analysis is more efficient.
These values are shown in column $g_1$ of Table \ref{jointtab}.
As a consistency check we show in column $g_2$ the factor gained
by the joint sample
when compared with the most appropriate case of the published WMAP results.
The largest gain is on the cluster abundance scaling, $\sigma_8\,\Omega_m^{0.5}$.
As we saw, this parameter is roughly the first principal component of the
cosmic shear and its error is well determined by cosmic shear alone.
So, what makes more sense here is to compute the gain with respect to
the cosmic shear sample and not to the CMB sample. In this case  we find 
the joint sample brings no gain $(g_1=1.1)$.

\begin{figure*}
\centering
\includegraphics [height=13cm] {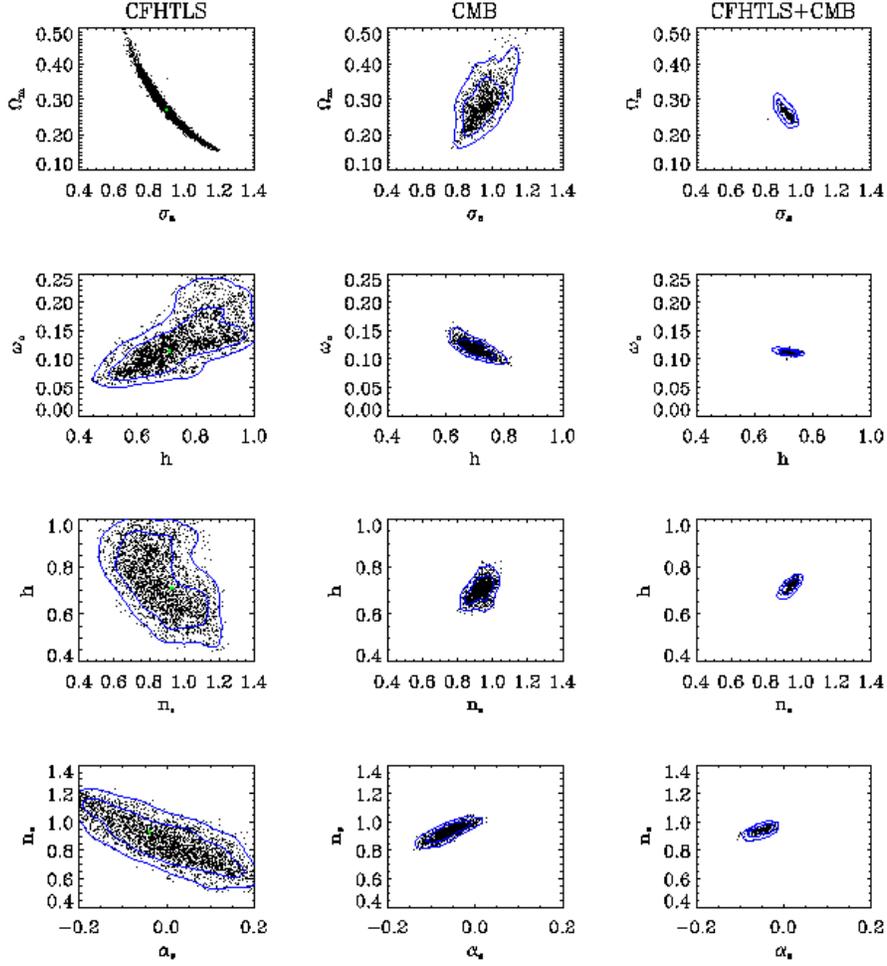} 
\caption{Marginalised 2-dimensional $68\%$ and $95\%$ contours
from the three samples. These are the most relevant
plots to illustrate the largest gains on the parameters precisions.} 
\label{joint2d}
\end{figure*}

Keeping in mind that the
combined result of 2 independent experiments with 
errors of the same order has already a gain of $\sqrt{2}$, 
we consider the combined analysis to be efficient for a
certain parameter if 
$g>\sqrt{2}$. Hence, the efficiency is higher for
the dark matter density, the hubble parameter, $\sigma_8$
and the spectral indexes. To illustrate this result, we plot
in Fig. \ref{joint2d} the pairs of parameters where the orthogonality 
between CMB and CFHTLS contours is most striking,
among all possible pairs. 
These are contours of equal likelihood, containing $68\%$ and $95\%$
of the sample.
All the 4 cases involve only the efficient parameters.
Furthermore, they correspond to cosmic shear $Y_1$ and $Y_2$ related
well constrained cases we found in the previous section. Thus, we found that
projections of the best constrained cosmic shear principal components
are orthogonal to the corresponding CMB contours, which shows a complementarity
between cosmic shear and CMB.
\begin{figure*}
\centering
\includegraphics [height=8cm] {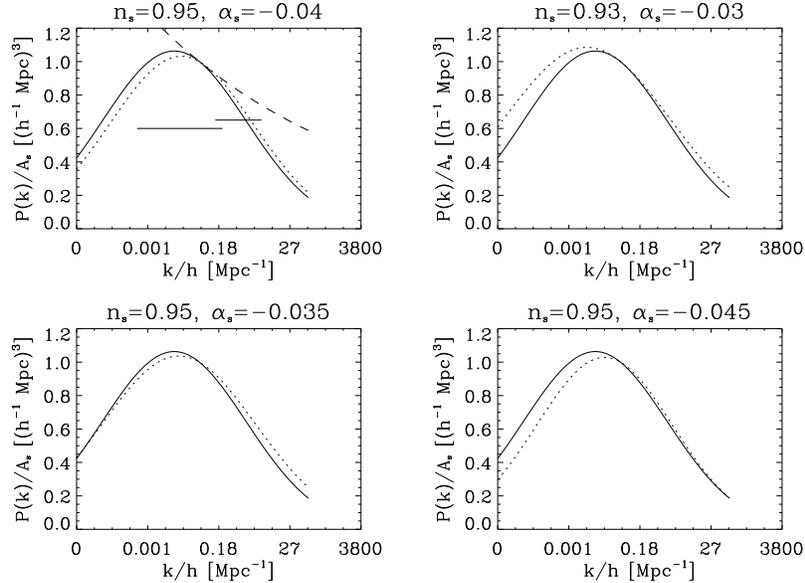} 
\caption{Primordial power spectrum parameterized by $n_s$ and $\alpha_s$.
The values written 
  in the panel titles define the dotted lines, while, in each panel, 
the solid lines are plotted with the fiducial values
$(n_s=0.93 \,\alpha_s=-0.04)$.
The scale ranges probed by the CMB and the cosmic shear are shown in the upper
left panel.
}
\label{nsas}
\end{figure*}

To understand the origin of some of this complementarity, let us consider
the $n_s$/$\alpha_s$ case. The gain on both parameters is around 2,
even though, as we saw, the cosmic shear by itself is not very sensitive to
the running spectral index. In Fig. \ref{nsas} we plot the 
primordial power spectrum from Eq. (\ref{pkpri}) as function of the linear 
wavenumber. The spectral indexes parameterize the shape of the spectrum
and have no further role on its evolution. Thus the opposite 
behaviour of the CMB/cosmic shear responses to a change on the indexes 
may be understood from these plots. The solid line in all panels is the
fiducial model $(n_s=0.93\,,\alpha_s=-0.04)$. It bends away from a power
law ($\alpha_s=0$, the dashed line on the upper left panel) from the
pivot point. The dotted lines are deviations from the fiducial model,
they correspond to the indexes values written as the panels titles.
In the top panels one parameter is changed at a time. On the left,
a change on $n_s$ produces changes of opposite signs at both ends of
the spectrum. On the right, changing $\alpha_s$ raises both ends of
the spectrum. The bottom panels show how it is possible to mimic the
fiducial spectrum for large (small) scales by changing both indexes in the
same (opposite) direction. 

On the first panel, the solid horizontal lines show the scale ranges probed
by the CMB (the line on the left) and the cosmic shear (on the right)
data used in this work.
These intervals were found by using the calculations of \cite{TEGZAL02},
in particular their fitting $k\approx \sqrt(3'/\theta)\,[h/Mpc]$. 
Hence, the 2 bottom plots lead to expect an upper left - lower right
direction of degeneracy for the cosmic shear and an orthogonal one
for the CMB.

The shape of the $n_s/h$ lensing degeneracy has a similar origin. The slope of
the power spectrum at the scales probed by the cosmic shear is $\Omega_m \,h$.
A raise of $h$, increases the power at small scales that must be compensated by
a decrease in $n_s$.

\begin{figure}[h]
\centering
\includegraphics [height=7cm] {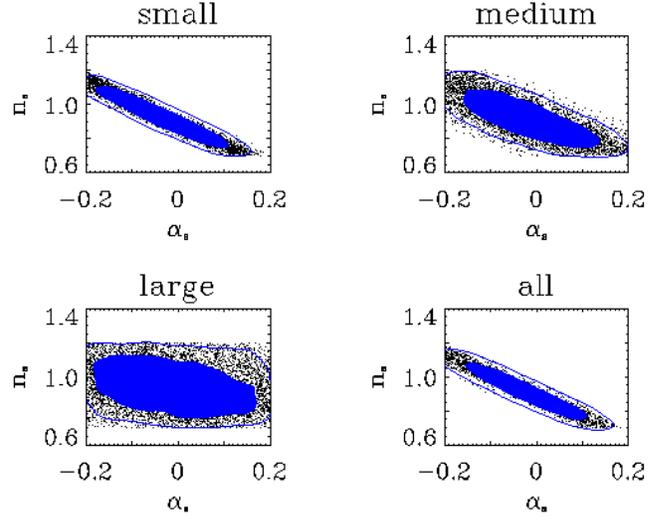} 
\caption{$n_s\,\alpha_s$ MCMC sample points and $68\%$ and $95\%$ confidence
levels. The models likelihoods were
evaluated using 4 different sets of angular scales - small, medium, large and
all, defined in the text. The "small" case is the one that better constrains the
$n_s\,-\alpha_s$ degeneracy found in the 7-dimensional analysis.} 
\label{smla}
\end{figure}

In order to have an explicitly view of the cosmic shear $n_s/\alpha_s$ degeneracy
scale dependence, we produced a new set of 4 cosmic shear MCMC chains.
This time,  we only allowed the scalar spectral indexes to change
and kept the other 5 of the 7 original parameters (see Table \ref{cslimit})
at the fiducial values.
The chains were built following the
procedure detailed in section 3. Due to the small number of parameters probed,
convergence was very rapid and a burn-in of 300 elements was enough to reach the
equilibrium. The models shear dispersion and likelihood were computed for 
4 cases, distinct on the angular scales used. These are (in arc minutes):
\begin{itemize}
\item
"small" scales: $(0.6,1,1.3,2,3,4,5,6)$
\item
"medium" scales: $(8,10,15,20,25,30)$
\item
"large" scales: $(40,50,60,80,100,120)$
\item
"all": all the above 20 angular scale points, which are the ones used in the
7-dimensional MCMC cosmic shear chains. 
\end{itemize}  
Figure \ref{smla} plots the sample points and contours obtained.
It shows that even inside the comparatively small region of Fig. \ref{nsas}
probed by the cosmic shear, the scale dependency of the $n_s/\alpha_s$
degeneracy constraint is detected, with the best constraint coming from the
smallest scales.   
The signal from the largest of the cosmic shear scales is more dispersed as
it is shown at the bottom left panel of Fig. \ref{smla}. Notice, in this case,
the orientation of the contours is cut by the MCMC exploration limits imposed.

\begin{figure}[h]
\centering
\includegraphics [height=3cm] {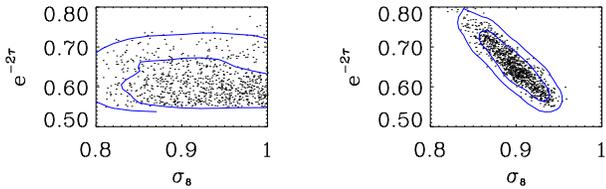} 
\caption{Marginalised 2-dimensional $68\%$ and $95\%$ contours.The first plot
is from the CMB sample and
the second one from the joint sample.} 
\label{tausig8}
\end{figure}

There are parameters for which no gain was found, for instance,
the measurement of
$\omega_b$ is dominated by the CMB. For $\tau$, even though the
cosmic shear is not sensitive to it, the introduction of cosmic shear data
strengthens the $\sigma_8-\tau$ correlation allowing to lower the errors on
$\sigma_8\,e^{-2\,\tau}$ (Table \ref{jointtab}). Thus, even though we do not
predict a gain on the measurement of $\tau$ from CFHTLS+CMB data,
future cosmic shear surveys, through a more precise measure of $\sigma_8$,
will be helpful in its determination.
In Fig. \ref{tausig8} its is shown the correlation between $\sigma_8$
and $e^{-2\,\tau}$, the factor by which the CMB at small scales
is damped after reionization. The information provided by the cosmic shear
is clear.

\section{Discussion}

We studied the determination of cosmological parameters by a 
CFHTLS-Wide type of experiment. For this we have made some assumptions 
that may not exactly match the real situation.
First of all, the real survey features will not be identical to the 
used ones. In particular, we made the optimistic assumption that all the
$170\,\rm{deg}^2$ of the survey area will be usable, whereas 
a size of $130\,\rm{deg}^2$ would be more realistic\footnote{This is 
 mainly due to the masking process that reduces by about 20\% the total 
   sky coverage of deep surveys}.
Furthermore, since we assumed the redshift of the sources perfectly known, 
there will be an extra source of error coming from the marginalisation over the 
real sources redshift distribution. The same happens with the marginalisation
over other cosmological parameters not taken into account in this study, such as
the equation of state of dark energy or the neutrinos density.
The calculations into deep non-linear regime using the {\sc halofit}
formula may also bring an extra source of errors.
 
We need also to keep in mind that we made a systematics-free study, with data
errors being dominated by Poisson noise and cosmic variance.
B-mode contamination due to systematic
residuals could significantly
increase uncertainty in the noise correction process and on the
way it should be handled in the covariance matrix.
However, it has recently been shown that the residual 
B-mode on the {\sc Virmos}-Descart data could be further corrected and 
eventually set to zero.
This result follows from a better understanding 
of the PSF variation across the CCD chips (see \cite{HOEKSTRA04}
and \cite{VWMH04})
and it is expected that the CFHTLS will be at least as good as {\sc 
Virmos}-Descart in terms of image quality, hence we set B-modes
to zero in this work. 

On the other hand, our exploration of cosmological parameters 
  is more realistic and 
 accurate than previous studies. We use a non-diagonal data covariance matrix, allowing for
correlation between scales. Furthermore the MCMC method probes the real shape of
the posterior PDF, without assuming a gaussian distribution. 
 It is also important to stress that  the 2-points correlation functions 
  do not contain all
the cosmic shear information. In particular, we did not use 
higher order statistics, like the already detected convergence skewness
(\cite{PENetal03}) or the three point shear correlation
functions (\cite{BERNMVW02}), that may help also to break
degeneracies (\cite{BERNVWM97}).  

We explored the cosmological parameters space using cosmic shear,
describing the results in the context of a principal
components analysis, and found a set of parameters degeneracies 
orthogonal to CMB ones. This led to predict a gain of the order of 2 or 3 for
several parameters, when combining CFHTLS-Wide data with WMAP and CBI data. 
This means, for example, precisions of $\sigma(\sigma_8)=0.03$ and 
$\sigma(\Omega_m)=0.02$.
This result is consistent with the parameters determinations of 
\cite{CONTALDHOEK03} that combines CMB data with the Red-Sequence Cluster Survey
(RCS) data, where they found $\sigma(\sigma_8)=0.05$ and 
$\sigma(\Omega_m)=0.03$.
From a quick Fisher matrix calculation, the ratio between
parameters errors determined by 2 cosmic shear experiments is approximately
given by a ratio of the experiments survey areas (A), galaxy densities (n) and
intrinsic ellipticity dispersions ($\sigma$)  (see for example \cite{HUTEG99}):
\begin{equation}
\label{lensratio}
g_{cs}=\left({{A_1\,n_1^2\,\sigma_2^4}\over
{A_2\,n_2^2\,\sigma_1^4}}\right)^{1\over2}.
\end{equation}
The ratio
between CFHTLS-Wide and RCS (with a size $A=53 \,\rm{deg}^2$) is 1.8.

As compared to the fiducial reference survey used by \cite{ISHAK03},
this simple estimation predicts the precisions obtained with their reference
survey should be 4
times better than the ones from our configuration. But we find the same, or only
slightly bigger, $\sigma$ values.
First of all, Eq. (\ref{lensratio}) gives only an upper limit for the combined gain.
In fact, for a parameter whose determination is dominated by CMB, the 
cosmic shear contribution to the joint result must be smaller than this value.
The main reason for this discrepancy is, however, the
fact that our 
degraded configuration, as compared to the survey in \cite{ISHAK03},
is partly compensated by the inclusion of smaller angular scales.
In fact, we saw that it is in the non-linear regime that lies, not only the
cosmic shear greatest sensitivity to the cosmological parameters, but also the
cause of its orthogonality to CMB.

\section{Conclusions}

\begin{figure}
\centering
\includegraphics [height=6cm] {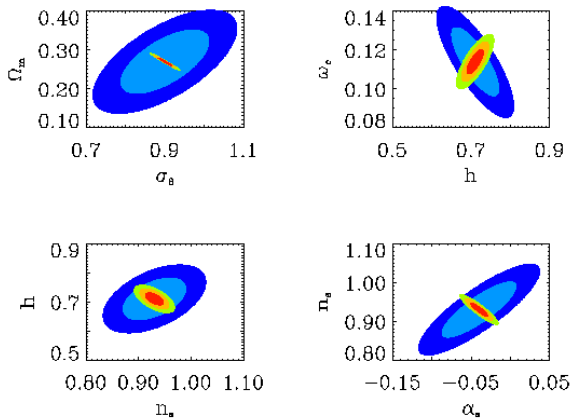} 
\caption{68\% and 95\% C.L. for the 4 most orthogonal cases found in
section 5. Blue is WMAP+CBI data and red shows predictions for
the wide field space telescope - cosmic shear parameters of Table (\ref{snaptab})
 (99 \% C.L. is also shown in this case).} 
\label{snapcmb}
\end{figure}

The CMB/cosmic shear complementarity opens
good prospectives for the determinations of cosmological parameters
by combining CMB and cosmic shear data sets. In fact, even for CFHTLS, whose
contours are, in general, noticeably larger than the
WMAP+CBI ones (Fig. \ref{joint2d}), we
predict non neglectable gains. Figure \ref{snapcmb} shows what can be
expected with future space telescope data.
\begin{table}
\caption{Cosmic shear : Wide field space telescope illustration specifications.}
\label{snaptab}
\begin{center}
\begin{tabular}{ll}
\hline \hline
 Size of the survey: &A=$1000\,{\rm deg} ^2$ \\
 Density of galaxies: & $n_g=50\,{\rm arcmin}^{-2}$ \\
 Intrinsic ellipticity dispersion:& $\sigma_\epsilon=0.3$ \\
 Scales probed:& $0.6\,'<\theta<5\deg$\\
 \ & $40<l<18000$\\
 \hline \hline
\end{tabular}
\end{center}
\end{table}
These results were produced with a cosmic shear Fisher matrix analysis,
using Eq. (\ref{fisher}) and the fiducial model of Table \ref{fiduc} (except for
the redshift of the sources which was moved to $z_s=1.1$).
For this illustration, the data covariance matrix
of Eq. (\ref{bigcov}) was computed for the configuration shown in
Table \ref{snaptab}, which is close to the
SuperNova Accelerator Probe / Joint Dark Energy Mission (SNAP/JDEM)
"Wide+" case of \cite{REFSNAP3}. The CMB ellipses are plotted from
the parameters covariance matrix found with our WMAP+CBI chains.

In summary, we found the best constrained parameters by 2-point cosmic shear
correlation functions to be $\sigma_8\,\Omega_m^{0.5}$ and 
$\Gamma\,n_s^{0.6}$ (with $\Gamma=\Omega_m\,h$).
We have shown that 2-dimensional degeneracies defined by these
parameters plus another one defined by $n_s$ and $\alpha_s$ are orthogonal
to CMB degeneracies. 
Due to this CMB/cosmic shear complementarity, current weak lensing surveys,
such  as
the CFHTLS, already have the potential to improve the precision on several
cosmological parameters. In particular, a better knowledge of $\alpha_s$ will
have an impact on inflationary scenarios. The crucial information 
 provided by the
cosmic shear comes from the small scales it probes. Thus, it provides an
additional possibility, along with galaxy redshift surveys 
and Lyman-$\alpha$ forest, to combine with CMB data. 

\acknowledgements{We thank D. Bond, F. Bernardeau, S. Prunet, C. Contaldi, 
D. Pogosyan, K. Benabed and R. Gavazzi for useful discussions.
We thank CITA for the use of the DOLPHIN cluster, where the
chain calculations were performed, and
the TERAPIX data center for additional computing facilities. 
IT is supported by a Funda\c{c}\~ao para a Ci\^encia e a Tecnologia (FCT)
scholarship.
}

\end{document}